\documentclass[a4paper]{jpconf}
\usepackage{graphicx}
\begin{document}
\title{In-situ Surface Contamination Removal and Cool-down Process of the DEAP-3600 Experiment.}

\author{Pietro Giampa, for the DEAP collaboration}

\address{Queen’s University, Physics Dept., Kingston, Ontario K7L 3N6, Canada}

\ead{pietro@owl.phy.queensu.ca}

\begin{abstract}
The DEAP-3600 experiment is a single-phase detector that uses 3600 Kg of liquid argon to search for Dark Matter at SNOLAB, Sudbury, Canada, 6800 ft. underground.
The projected sensitivity to the spin-independent WIMP-nucleon cross-section is $10^{-46}$ cm$^{2}$ for a WIMP mass of 100 GeV. \\
A key experimental requirement is the reduction of any possible source of background that would mimic a Dark Matter signal. 
This document will review how radiogenic surface backgrounds were reduced in-situ by removing 500 microns of acrylic from the innermost part of the detector with a resurfacing robot. Furthermore it will review the transient cool-down process of the experiment, necessary to reach cryogenic operating temperature.

\end{abstract}

\section{DEAP-3600 Experiment}
\begin{figure}
\begin{center}
\includegraphics[width=70mm]{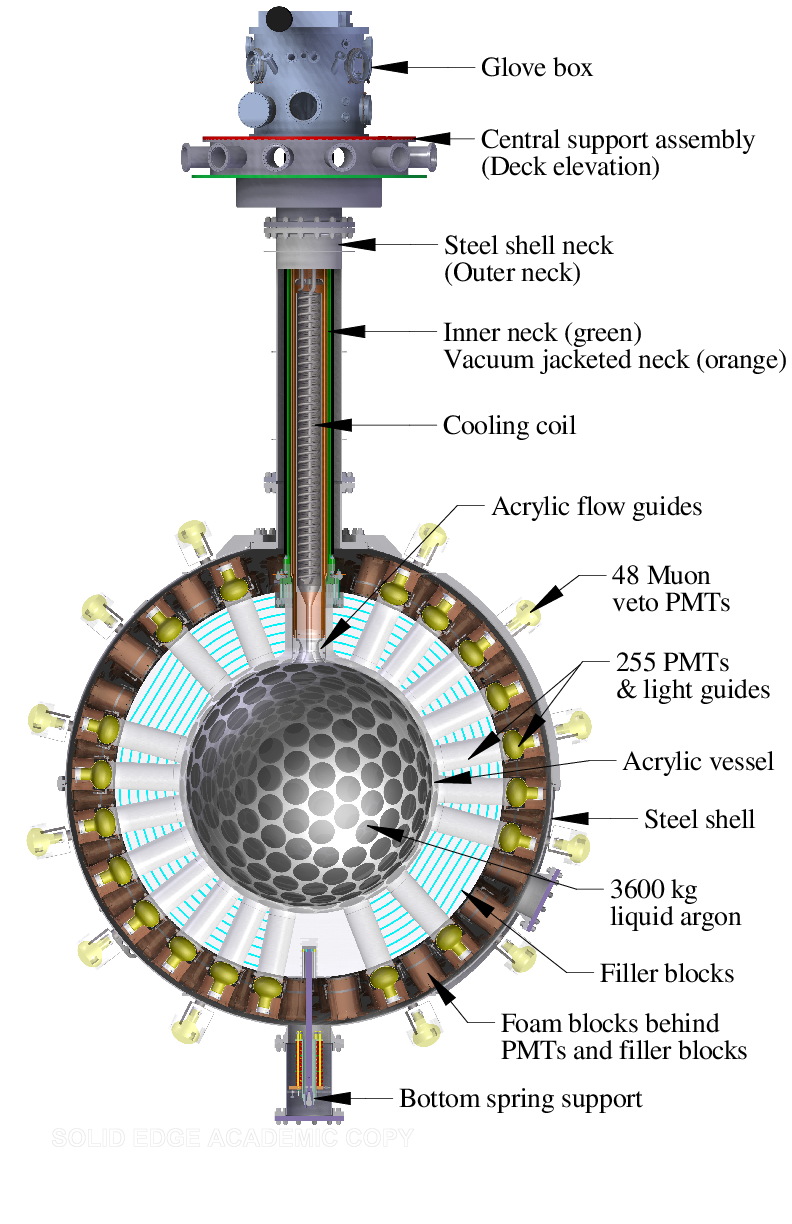}
\caption{The DEAP-3600 detector. 3600 kg of liquid argon are contained within an acrylic vessel (AV) of 85 cm radius. Scintillation light is collected by 255 8'' Hamamatsu R5912 HQE photomultipliers, coupled to the AV with 50 cm long acrylic light guides.}
\label{fig:DEAP_Eng}
\end{center}
\end{figure}

DEAP-3600 is a 3600 Kg single-phase liquid argon experiment, located 6800 feet underground at the SNOLAB facility in Sudbury, Canada. Designed to detect weakly interacting massive particles (WIMPs), this detector has a projected spin-independent WIMP-nucleon cross-section sensitivity of $10^{-46}$ cm$^{2}$ for a WIMP mass of 100 GeV \cite{DEAPmarcin}.
To describe the structure of the experiment it is necessary to start from its core (Fig.\ref{fig:DEAP_Eng}).
The target material is contained within a spherical acrylic vessel (AV) 85 cm in diameter. 
When a charged particle interacts with liquid argon, it generates ultraviolet scintillation light (128 nm). A layer of wavelength shifter, 1,1,4,4-tetraphenyl-1,3-butadiene (TPB) \cite{DEAPtpb}, was deposited on the inner surface of the AV to convert the scintillation light into the visible range. This light signal is collected by 255 8'' Hamamatsu R5912 HQE Photomultipliers \cite{DEAPmark}, optically coupled to the AV via acrylic light-guides (LG) (Fig.\ref{fig:DEAP_Eng}). 
In addition, the LGs also act as shielding from external backgrounds and the radioactivity of the PMT glass. Both the AV and LG are covered with specular and diffusive reflectors to prevent light loss. The space between LGs was filled with polyethylene filler blocks, which provide both thermal and external background shielding. The AV is contained within a stainless steel outer vessel, at the center of a 402 m$^{3}$ water shield tank. 
Instrumented on the outer vessel are outward looking PMTs that function as Cherenkov veto for cosmogenic muons. Since February 2015, following the data acquisition (DAQ) system commissioning, much effort was dedicated to the calibration of the PMTs and detector optical response. Multiple optical sources have been used, including an isotropic light source deployed at the center of the experiment and a separate system of light injection fibers. 
The detector construction was completed in August 2015, and the experiment is currently in the transient cool-down process. DEAP-3600 is expected to start a first physics data taking run by the end of 2015.

\section{The Resurfacer}
\subsection{Motivation}
\begin{figure}
\begin{center}
\includegraphics[width=80mm]{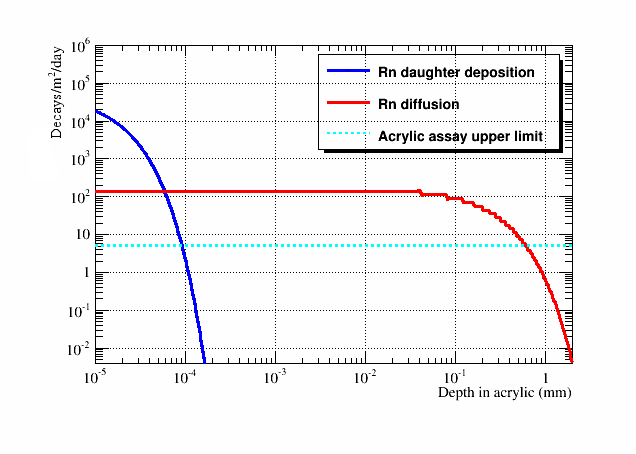}
\caption{Rn contamination in the AV as a function of acrylic depth from the Rn daughters deposition and Rn diffusion. Also shown the upper limit on the AV bulk acrylic from direct assay measurement.}
\label{fig:DEAP_Res_Mot}
\end{center}
\end{figure}

Radioactive contamination on the surface of the detector can significantly reduce the sensitivity of the experiment \cite{DEAP1bkg}, and it is necessary to ensure that this type of background is limited to a contribution of less than 0.2 events per 3 years. 
Particular attention was given to the selection and production of the acrylic used for the AV, specially produced in a Rn reduced environment. Direct assays of the acrylic set an upper limit on its activity to 10$^{-19}$ g/g of $^{210}$Po. While methodical care was taken at every stage of construction, certain installation tasks required the AV to be exposed to SNOLAB lab air and therefore Rn. 
Such exposure can lead to two contamination processes: Rn daughter’s deposition on the surface of the acrylic and Rn diffusion into the bulk acrylic \cite{Wojcik}. These contaminations can be reduced by removing a thin layer of acrylic from the inner most portion of the AV. The expected contamination level as a function of acrylic depth is shown in Fig.\ref{fig:DEAP_Res_Mot}.

\subsection{Design}
\begin{figure} 
\begin{center}
\includegraphics[width=70mm]{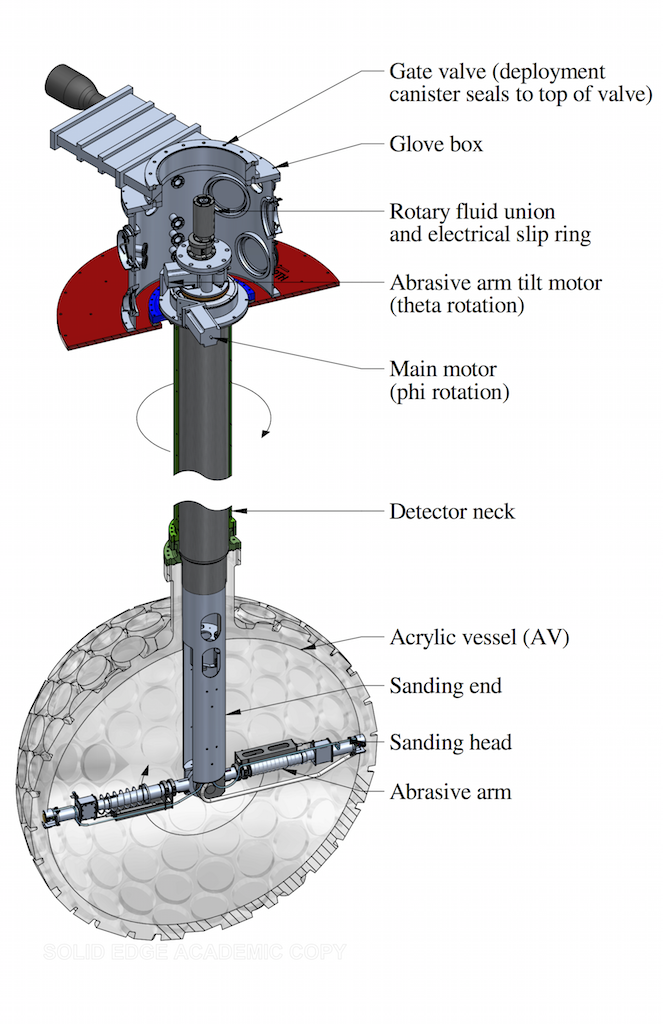}
\caption{Engineering representation of the Resurfacer deployed inside the DEAP-3600 acrylic vessel. The two sanding arms are capable of uniformly reaching the entire inner surface of the experiment, and are mounted with position sensors that can accurately measure the amount of removed acrylic.}
\label{fig:DEAP_Res}
\end{center}
\end{figure}

The Resurfacer features two retractable sanding arms, one for each AV hemisphere, which are deployed at the center of the AV via an 18 foot long deployment tube (Fig.\ref{fig:DEAP_Res}).
These arms are designed with a series of springs which are used to apply constant force on the acrylic, hence obtaining uniform acrylic sanding across the AV inner surface. 
The sanding action is performed by the heads of the sanding arms, which are mounted with 2'' sanding pads connected to a spinning motor located in a water-resistent enclosure inside the arm itself.  The efficiencies of each sanding arm was individually measured ex-situ with multiple tests, and they were estimated to remove acrylic at a rate of 9 g/hr.
The Resurfacer also features two fluid lines that bring ultra-purified water (UPW) directly to each sanding head, and one return line that extracts the mixture of UPW and sanded acrylic. 
The sanding arms reach the entire AV inner surface by moving in a spiral motion. This motion is controlled by two motors located at the top of the deployment tube and connected to the arms via stainless steel mechanical connections. 
Additionally, all materials used for the Resurfacer were carefully selected and subjected to Rn emanation assay to ensure that the experiment backgrounds budget was not exceeded.

\subsection{Operation}
Once the Resurfacer was deployed inside the AV, the detector was sealed and maintained in a low radioactive environment. This was achieved by keeping a 3 psig positive pressure with ultra-purified nitrogen gas, obtained by filtering boil-off nitrogen gas through a 75 g active-charcoal cold trap \cite{n2}. 
Conditions were preserved even during the extraction phase, thanks to a series of transfer volumes, also kept under ultra-purified nitrogen gas purge, and the robot was removed without exposing the inner surface of the AV to lab air.
In this configuration the Resurfacer was successfully run for over 200 hours and removed 500 microns of acrylic from the inner surface of the AV. 

\section{Cool-down Process}
The crucial task of the cool-down process is to bring the experiment from room temperature (293 K) to cryogenic temperature (87 K) filled with liquid argon in the fastest and safest mode possible. To accomplish this the cool-down rate needs to be such that thermal induced stress from local and time variations in temperature within the acrylic are reduced and argon freezing is avoided. 
There are two physical parameters that can be used to control the cool-down rate: the heat load generated by the cooling element (cooling coil) and the pressure inside the vessel.
In order to determine the optimal cooling rate it was necessary to study and understand all possible forms of heat transfer, which for DEAP-3600 are the following three: heat transfer between the AV surface and the Ar gas, heat transfer throughout the acrylic, and heat transfer between the cooling coil and the Ar gas. 
The last two have been carefully studied during the commissioning of the cooling coil at SNOLAB in May 2014 and with different ex-situ stress tests completed on acrylic blocks of the same material used for the AV. 
The heat transfer between the AV surface and the Ar gas was instead extrapolated from empirical values.
The experiment will be cooled at low pressure of 9 psia, reached with small increments of 0.2 psi, with a cooling power from the cooling coil that will never exceed 1 kW. This process is expected to take up to 2 weeks of operation.

\section{Conclusions}
The DEAP-3600 experiment has completed construction and it is expected to start its physics program by the end of 2015. Calibration data has been collected since early 2015, including the use of radioactive sources and the deployment of a uniform light source at the center of the detector. 
A resurfacing robot, the Resurfacer, was specifically designed and assembled to reduce potential surface contaminations introduced during construction. The Resurfacer was deployed in October 2014 inside the AV, and it ran for over 200 hours removing 500 microns of acrylic from the AV inner surface.
In order to achieve a high safety factor DEAP-3600 will go through a low pressure cooling process. The detector will be brought from room temperature to cryogenic temperature at a low pressure of 9 psia with a cooling power from the cooling coil that will never exceed 1 kW. All required hardware has been tested and commissioned in-situ in early 2014.

\end{document}